\documentclass{iaus}
\usepackage{graphicx}

\newcommand{\Dnu}[1]{\Delta \nu_{#1}}
\newcommand{\dnu}[1]{\delta \nu_{#1}}
\newcommand{\half}{{\textstyle\frac{1}{2}}}
\newcommand{\sixth}{{\textstyle\frac{1}{6}}}
\newcommand{\tenth}{{\textstyle\frac{1}{10}}}

\title[Asteroseismology, a tool for transit studies] 
{Measurements of Stellar Properties through \\ Asteroseismology:
\\ A Tool for Planet Transit Studies}

\author[Kjeldsen, Bedding \& Christensen-Dalsgaard]   
{Hans Kjeldsen$^1$, Timothy R. Bedding$^2$ \and \\
J{\o}rgen Christensen-Dalsgaard$^1$}

\affiliation{$^1$Danish AsteroSeismology Centre, Department of Physics and Astronomy, University of Aarhus, Ny Munkegade, DK-8000 Aarhus C, Denmark, email: {\tt hans@phys.au.dk} and {\tt jcd@phys.au.dk}
\\ $^2$Institute of Astronomy, School of Physics A28, University of Sydney, NSW 2006, Australia.
email: {\tt bedding@physics.usyd.edu.au} }

\pubyear{2008}
\volume{253}  
\pagerange{1 -- 10}
\jname{Transiting Planets}
\editors{}
\begin{document}

\maketitle

\begin{abstract}
Oscillations occur in stars of most masses and essentially all stages of
evolution.  Asteroseismology is the study of the frequencies and other
properties of stellar oscillations, from which we can extract fundamental
parameters such as density, mass, radius, age and rotation period.  We
present an overview of asteroseismic analysis methods, focusing on how
this technique may be used as a tool to measure stellar properties 
relevant to planet transit studies. We also discuss details of the Kepler
Asteroseismic Investigation -- the use of asteroseismology on the Kepler
mission in order to measure basic stellar parameters.  We estimate that
applying asteroseismology to stars observed by Kepler will allow the
determination of stellar mean densities to an accuracy of 1\%, radii to
2--3\%, masses to 5\%, and ages to 5--10\% of the main-sequence lifetime.
For rotating stars, the angle of inclination can also be determined.

\end{abstract}

\firstsection 
\section{Introduction}

Asteroseismology -- the study of stellar oscillations -- is a relatively
new and growing research field in astrophysics.  The analysis of
frequencies and other properties of stellar oscillations allows us to
constrain fundamental parameters of stars such as density, mass, radius,
age, rotation period and chemical composition.

Oscillations are found in stars of most masses and essentially all stages
of evolution.  The amplitudes and phases are controlled by the energetics
and dynamics of the near-surface layers.  The frequencies are determined by
the internal sound-speed and density structure of the star, as well as
rotation and (in some cases) magnetic fields.  Observationally, the
frequencies can be determined with exceedingly high accuracy compared to
any other quantity relevant to the internal properties of the stars.
Analysis of the observed frequencies, including comparison with computed
stellar models, allows determination of the properties of the stellar
interiors and tests of the physics used in the model computation, applied
under extreme conditions that cannot be matched in terrestrial
laboratories.

Rotation induces fine structure in the frequency spectrum, in the form of
rotational splitting.  The observed frequencies are determined by averages
over the stellar interior of the rotation rate, which in general varies
with position within the star. By comparing with independent determinations
of the surface rotation rate, or from rotational splittings for a
sufficient broad variety of modes, information about this variation can be
obtained.

Some of the basic parameters that may be obtained from asteroseismology are
crucial for understanding the fundamental properties of exoplanet systems.
Hence, asteroseismology will potentially be an excellent tool for
characterizing planet transit systems and, specifically, it will be able to
measure stellar radii to a relative accuracy of 2--3\%.

\begin{figure}[b]
\begin{center}
 \includegraphics[width=5.4in]{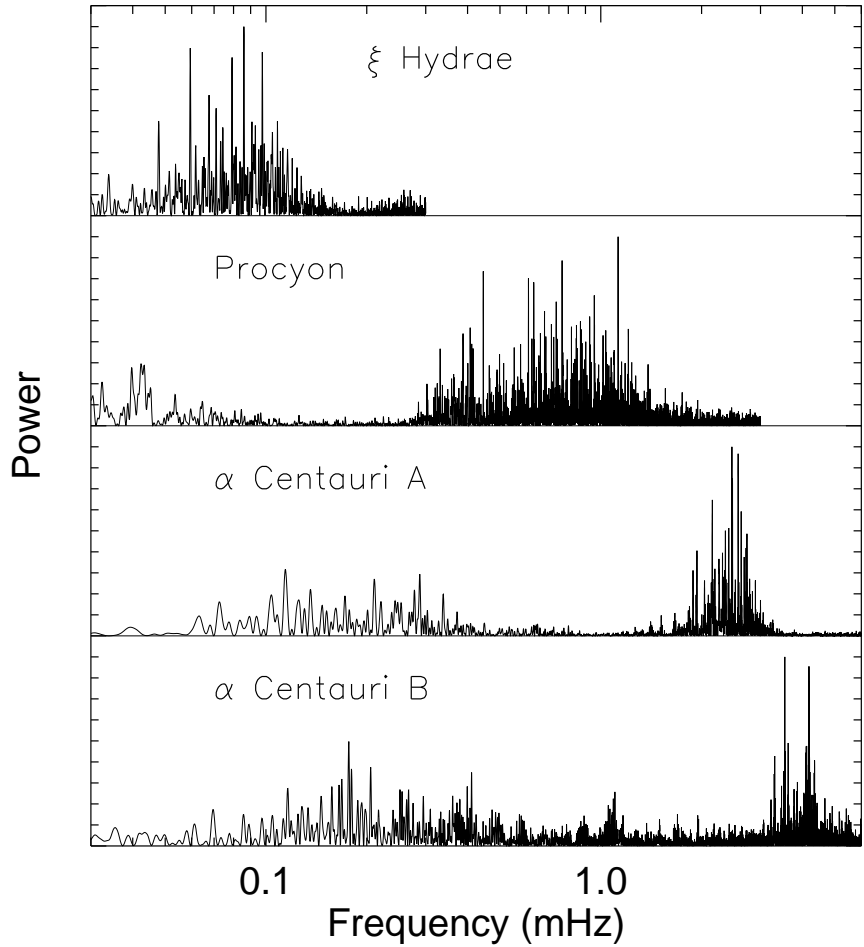} 
 \caption{\label{fig:fourstars} Power spectra of time series observed in
 radial velocity for the four stars $\xi$ Hydrae (G7\,III; \cite[Frandsen
 \etal\ 2002]{Frandsen02}), Procyon (F5\,IV; \cite[Arentoft \etal\
 2008]{Arentoft08}), $\alpha$~Centauri~A (G2\,V; \cite[Bedding \etal\
 2004]{Bedding04}) and $\alpha$~Centauri~B (K2\,V; \cite[Kjeldsen \etal\
 2005]{Kjeldsen05}).  The oscillation periods for the four stars are:
 around 3--4 hours for $\xi$ Hydrae, 20--25 minutes for Procyon, 7 minutes
 for $\alpha$~Cen~A and 4 minutes for $\alpha$~Cen~B.  These differences,
 and those in the detailed structure of the power spectra, reflect the
 differences in stellar properties (radius, mass, surface temperature and age).
 Note that the vertical scales are normalized -- the actual amplitudes
 decrease from top to bottom in the figure.  }
   \label{fig1}
\end{center}
\end{figure}

For a recent review on the present state of asteroseismology, see
\cite[Aerts \etal\ (2008)]{Aerts08} and \cite[Bedding \& Kjeldsen
(2008)]{Bedding08}. In Figure~\ref{fig:fourstars} we show examples of power
spectra for four different stars.  In each case, we can clearly see the
excess energy arising from stellar oscillations. The detailed properties of
those oscillations form the observational basis for the asteroseismic
analysis.

\begin{figure}[b]
\begin{center}
 \includegraphics[width=5.2in]{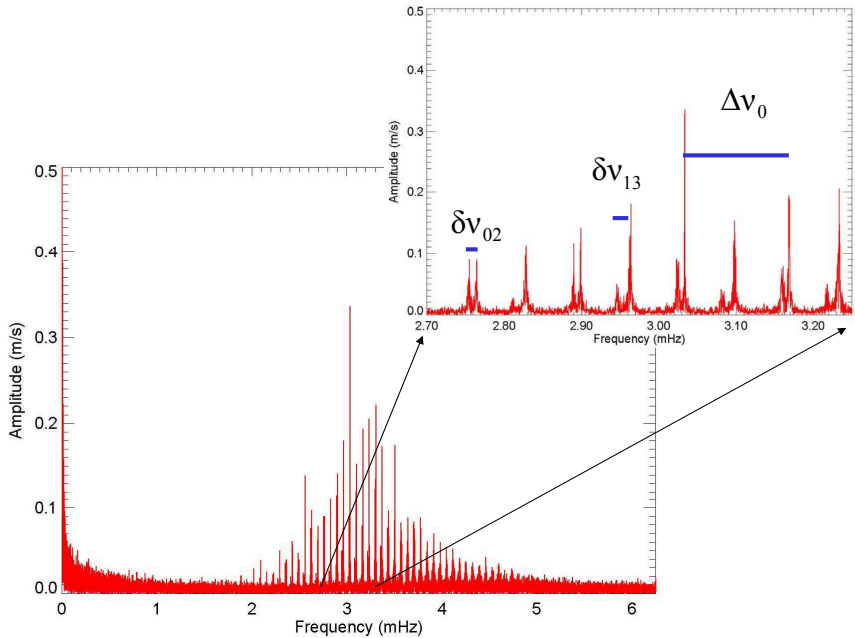} 
 \caption{\label{fig:golf} The power spectrum of oscillations in the Sun.
The data are full-disk radial velocity measurements obtained over 30 days
using the GOLF instrument on board the SoHO spacecraft (\cite[Ulrich \etal\
2000]{Ulrich00}, \cite[Garc{\'{\i}}a \etal\ 2005]{Garcia05}). The inset
shows the details of the p-mode structure, and we indicate the so-called
large and small separations, which contain information on the basic stellar
properties.}
   \label{fig2}
\end{center}
\end{figure}

\begin{figure}[b]
\begin{center}
 \includegraphics[width=5.2in]{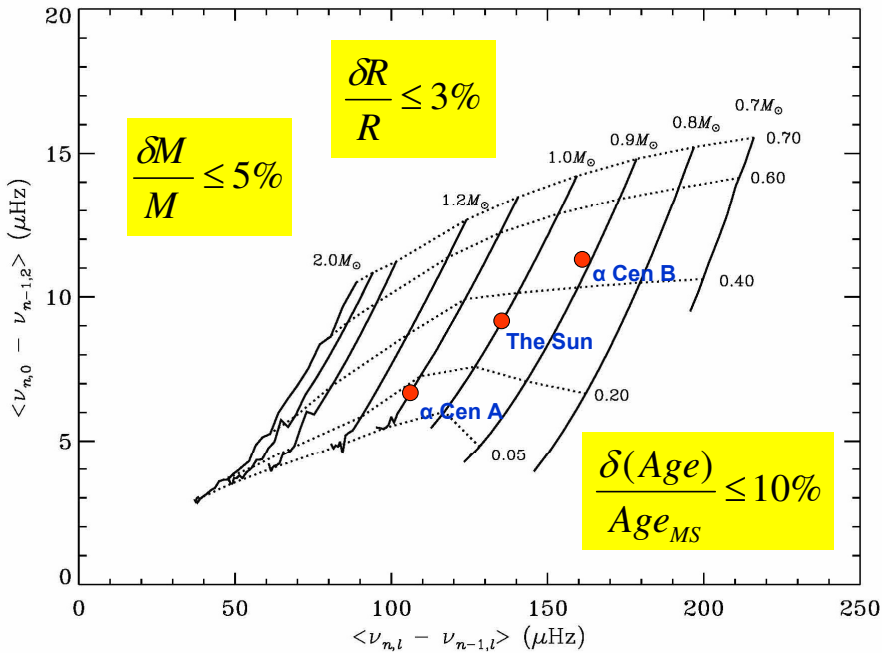} 
 \caption{\label{fig:CDdiagram} The so-called asteroseismic HR-diagram
 (\cite[Christensen-Dalsgaard 1993)]{Christensen-Dalsgaard93}, where the
 axes are the large and small frequency separations ($\Dnu{}$ and $\dnu{02}$).  The solid lines are
 evolutionary tracks (for fixed mass) and the dashed lines show constant
 core-hydrogen content, which is related to the age.  Based on simulations
 we have estimated the uncertainties for the measurements of mass, radius
 and stellar age using the seismic HR-diagram to be 5\% for the mass,
 2--3\% for the radius and 5--10\% for the age (relative to the total time
 on the main sequence). The mean densities can be measured to 1\%.  The
 positions of the Sun and $\alpha$ Centauri A and B are indicated, and it
 can be seen that $\alpha$~Cen~A is more evolved than the Sun while
 $\alpha$~Cen~B is less evolved, in agreement with traditional modeling.}
   \label{fig3}
\end{center}
\end{figure}

\begin{figure}[b]
\begin{center}
 \includegraphics[width=5.2in]{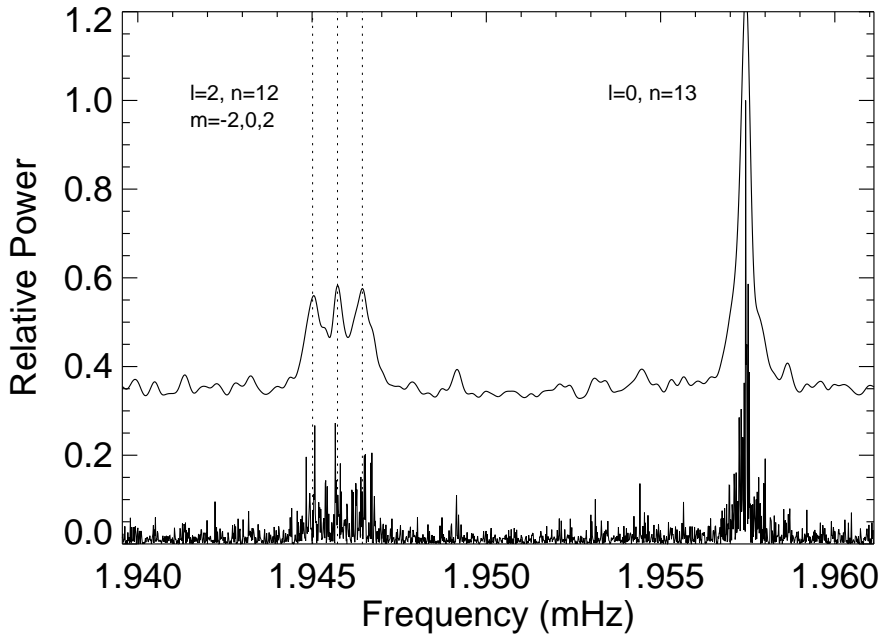} 
 \caption{\label{fig:rotation1}A close-up of the power spectrum of solar
 oscillations, from full-disk radial velocity measurements.  The data are
 based on 805 days of observing the Sun using the GOLF instrument on board
 the SoHO spacecraft (\cite[Ulrich \etal\ 2000]{Ulrich00},
 \cite[Garc{\'{\i}}a \etal\ 2005]{Garcia05}).  We show a small region of
 the power spectrum just below 2\,mHz. In this region one can see the power
 from modes with $l=2$, $n=12$ and $l=0$, $n=13$.  The $l=2$ multiplet has
 5 components ($m=0, \pm1, \pm2$) but, due to the inclination of the
 rotation axis of the Sun, only modes with $m=-2$, 0 and 2 are visible. The
 strength of the individual $m$-components can be used to measure the
 inclination axis of the stellar rotation.  Modes with $l=0$ have only 
 $m=0$ and are therefore not split by rotation.  The upper curve is a
 smoothed version of the power spectrum, shifted vertically for clarity.}
\end{center}
\end{figure}

\begin{figure}[b]
\begin{center}
 \includegraphics[width=5.2in]{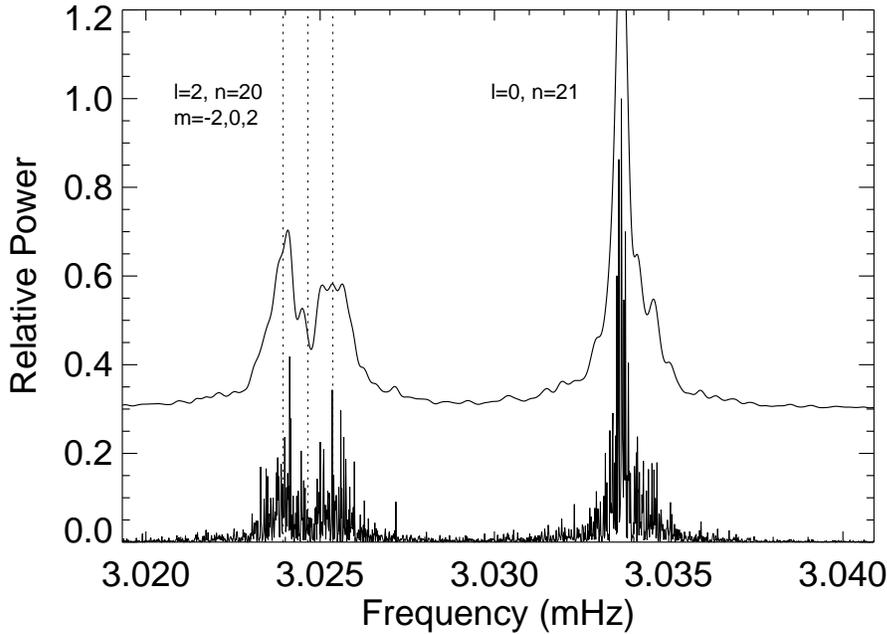} 
 \caption{\label{fig:rotation2} Same as Figure~\ref{fig:rotation1}, but
around a frequency of 3 mHz.  We see the multiplet with $n=20$ and $l=2$,
and the singlet with $n=21$ and $l=0$.  }
\end{center}
\end{figure}

\begin{figure}[b]
\begin{center}
 \includegraphics[width=5.2in]{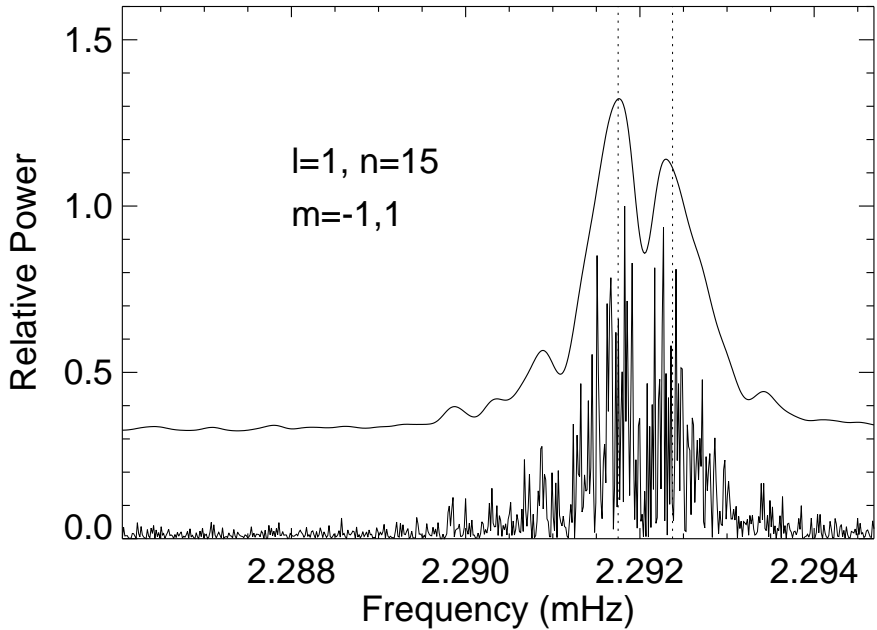} 
 \caption{\label{fig:rotation3} Same as Figure~\ref{fig:rotation1}, but at
a frequency just below 2.3\,mHz.  In this region one can see the power from
the triplet with $l=1$, $n=15$, $m=0, \pm1$.  Due to the inclination of the
rotation axis of the Sun, only modes with $m=-1$ and 1 are visible.  }
\end{center}
\end{figure}

\begin{figure}[b]
\begin{center}
 \includegraphics[width=4.8in]{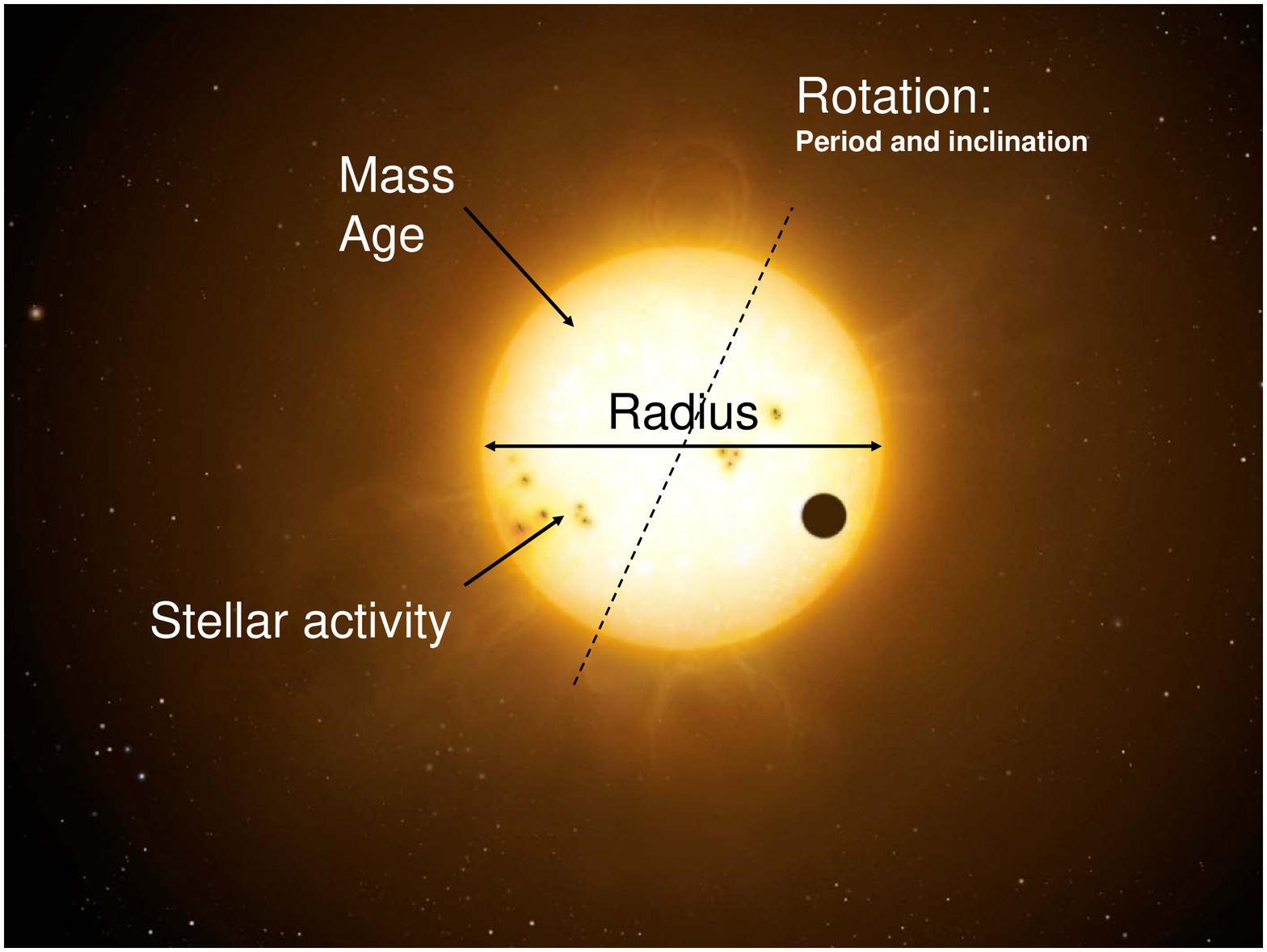} 
 \caption{Using asteroseismology we are able to measure detailed basic
 properties of stars that have a transiting planet. Asteroseismology is
 expected to become a crucial tool for the Kepler mission when measuring
 stellar parameters.} 
\end{center}
\end{figure}

\section{The relation between the stellar properties and the frequencies}

To obtain information about stellar properties and the underlying physics,
the observed oscillation frequencies must be compared with those computed
from models.  Major uncertainties affect the treatment of hydrodynamical
processes in stellar interiors. These include convective motions, with
possible overshoot into the surrounding convectively stable regions, and
flows and instabilities related to internal rotation.  Furthermore, the
evolution of rotation as a star evolves, with internal redistribution and
possibly surface loss of angular momentum, has potentially substantial --
and highly uncertain -- effects on the evolution of the star. Despite those
uncertainties in the knowledge of some of the detailed physics, we are able
to obtain quite detailed basic properties for a number of general stellar
parameters such as density and the amount of hydrogen in the core. 
The frequencies of stellar p-mode oscillations are related to the sound
travel time across the star. 
Since the sound speed,~$c$,  is given by
\begin{equation}
c^2  = \frac{\Gamma _{1} \, P}{\rho}  \simeq  \frac{5}{3}\frac{k_{B} T}{\mu m_{u}},
\end{equation}
(where the approximation assumes an ideal gas), we are basically measuring
the average ratio between pressure and density in the stellar interior.

The exact frequencies of stellar oscillations depend on the detailed
structure of the star and on the physical properties of the gas.  However,
one may use asymptotic theory to derive a simple relation for the mode
frequencies (see, for example, \cite[Christensen-Dalsgaard
2004]{Christensen-Dalsgaard04}).  Each oscillation mode is characterized by
three integers: the radial order~$n$, the angular degree~$l$ and the
azimuthal order~$m$.  The results of the asymptotic analysis for a
non-rotating star give the mode frequencies as
\begin{equation}
  \nu_{n,l} = \Dnu{} (n + \half l + \epsilon) - l(l+1) D_0.
        \label{eq.asymptotic}
\end{equation}
 Here, $\Dnu{}$ (the so-called large separation) depends on the average
stellar density, $D_0$ is sensitive to the sound speed near the core and
$\epsilon$ is sensitive to the surface layers.  It is conventional to
define $\dnu{02}$, the so-called small separation, as the frequency spacing
between adjacent modes with $l=0$ and $l=2$.  These separations are
indicated for the Sun in Figure~\ref{fig:golf}, together with the similar
quantity~$\dnu{13}$.  We can further define $\dnu{01}$ to be the amount by
which $l=1$ modes are offset from the midpoint between the $l=0$ modes on
either side. If the asymptotic relation holds exactly, then it follows that
$D_0 = \sixth\dnu{02} = \half\dnu{01} = \tenth\dnu{13}$.

In practice, the asymptotic relation does not hold exactly, even for the
Sun.  For example, the large separation depends on~$l$. However, we can
define average values for $\Dnu{}$ and $\dnu{02}$, and then can calculate how
those parameters depend on the stellar properties and the stellar
evolutionary stage.  An example of such a calculation is shown by the lines
in Figure~\ref{fig:CDdiagram}.  In this figure we also show the measured
position of the Sun, together with that of $\alpha$ Centauri A
(\cite[Bedding \etal\ 2004]{Bedding04}) and $\alpha$ Centauri~B
(\cite[Kjeldsen \etal\ 2005]{Kjeldsen05}).  This diagram confirms that
$\alpha$~Cen~A has a higher mass than the Sun and is more evolved,
while $\alpha$~Cen~B has a lower mass than the Sun and is less
evolved.

The exact positions of the tracks shown in Figure~\ref{fig:CDdiagram} will
depend on the detailed physical properties, including chemical composition.
However, the basic stellar parameters can still be estimated with some
robustness.  We have used simulations to estimate the accuracy with which
various parameters can be measured, assuming the stellar temperature is
already known to within 150--200\,K and the heavy-element abundance is
known to within a factor of two.  We find that
\begin{itemize}
\item stellar mean densities can be measured to 1\%,
\item stellar radii can be measured to 2--3\%,
\item stellar masses can be measured to 5\% and
\item stellar ages can be measured to 5--10\% of the main-sequence lifetime.
\end{itemize}

\section{Stellar Rotation}

Equation~\ref{eq.asymptotic} applies to a non-rotating star, for which the
frequency of each mode depends only on the radial order~$n$ and the angular
degree~$l$.  For a rotating star, the frequencies also depend on the
azimuthal degree,~$m$ (which takes on values of $m= 0, \pm 1, \ldots, \pm
l$).  Provided the rotation is slow, the frequencies are well approximated
by
\begin{equation}
  \nu_{n,l} = \Dnu{} (n + \half l + \epsilon) - l(l+1) D_0 + m  \Dnu{\rm ROT}.
        \label{eq.asymptotic2}
\end{equation}
Here, $\Dnu{\rm ROT}$ is related to the inverse of average of the internal
rotation period and we should therefore be able to infer stellar rotation
periods directly from the oscillation frequencies.

In principle, the inclination of the rotation axis can also be determined
because it affects the relative amplitudes of the different azimuthal
degrees.  This effect is discussed in detail by \cite[Gizon \& Solanki
(2003)]{Gizon03}.  For example, they show that for intensity observations
(oscillations detected in photometry) the dipole multiplets ($l=1$, $m=-1$,
0, 1) will have relative mode powers given by
\begin{eqnarray}
  P _{l=1, m=0} &= & \cos ^{2} i \; ,\\
  P _{l=1, m={\pm}1} &= & \half \sin ^{2} i \; .
\end{eqnarray}
 
In Figures~\ref{fig:rotation1},~\ref{fig:rotation2} and~\ref{fig:rotation3}
we show examples of rotational splitting for the Sun.  The splitting of the
$l$ = 1 and 2 modes is clearly seen.  The relative strengths of the different $m$
components is a result of the observed inclination of the solar rotation
axis. Since we observe the Sun in the equatorial plane, we have almost no
sensitivity to the modes with $l$=1, $m$=0 and those with $l$=2, $m$=$\pm
1$.  The structure of the power spectrum therefore indicates that the solar
interior is rotating in the surface equatorial plane of the Sun,
approximately corresponding to the plane of the ecliptic.

\section{The Kepler Asteroseismic Investigation}

The NASA Kepler mission (\cite[Borucki \etal\ 2008]{Borucki08}) will fly a
wide-field Schmidt camera with 0.95m aperture, staring at a single field
continuously for at least 3.5 years. Although the mission's principal aim
is to locate transiting extrasolar planets, it will provide an
unprecedented opportunity to make asteroseismic observations of a wide
variety of stars.  This will give the opportunity to measure global
properties and internal structure for a large number of stars across a
broad range of different types.  Plans are now being developed to exploit
this opportunity to the fullest. In particular, asteroseismic analysis of
Kepler data will provide an accurate determination of the radius for a
large fraction of the stars found to host planetary systems, as determined
from the transit analysis of the Kepler data, as well as estimates of the
ages of the systems.  Through investigation of a broad range of stars, the
asteroseismic investigation will also substantially improve our
understanding of general stellar evolution, and hence strengthen the use of
such modelling to further constrain the properties and evolution of the
stars and systems investigated in the Kepler extra-solar planet
programme. The purpose of the Kepler Asteroseismic Investigation (e.g.,
\cite[Christensen-Dalsgaard \etal\ 2007]{Christensen-Dalsgaard07}) is to
ensure that full use is made of this potential to benefit the Kepler
investigations of extra-solar planetary systems.

Kepler will give the possibility of observing at two cadences (1 minute and
30 minutes).  Due to the high frequency of the stellar oscillations Kepler
will not be able to make asteroseismic observations of all the 170,000
stars observed in long cadence mode (30 minute cadence).  We will therefore
need a target selection programme in order to optimize the asteroseismic
part of the Kepler programme. Based on simulations, we have estimated that
seismology can be done on unevolved main-sequence stars down to magnitude
$V=12.5$, and on evolved main-sequence stars (which have higher amplitudes)
down to $V=13.5$. For subgiant stars we may reach even fainter.  It should
be noted that Kepler will be the first mission where asteroseismology will
be applied to a large number of planet-hosting stars.

Finally, we note that accurate measurements of variations in the light
travel times for planets orbiting classical pulsating stars (the pulsations are here
used as accurate clocks) may provide a unique tool for the Kepler mission
to detect a number of additional planets.  The technique requires
accurate measurements of phase variations of the pulsations (better than
seconds) and the method will be most sensitive to long-period planets (1--2
years). For details on this technique we refer to \cite[Silvotti \etal\
(2007)]{Silvotti07}.

\end{document}